\documentstyle[12pt,epsfig]{article}
\def	\eqnum		#1{(\ref{#1})}       
	\newdimen\eqskip
	\newdimen\txtskip
	\baselineskip=\txtskip
\begin{document}

  \newcommand{\ccaption}[2]{
    \begin{center}
    \parbox{0.85\textwidth}{
      \caption[#1]{\small{{#2}}}
      }
    \end{center}
    }
\newcommand{\BS}{\bigskip}
\def\be{\begin{equation}}
\def\ee{\end{equation}}
\def\ba{\begin{eqnarray}}
\def\ea{\end{eqnarray}}
\def    \nn             {\nonumber}
\def    \=              {\;=\;}
\def    \frac           #1#2{{#1 \over #2}}
\def    \ret            {\\[\eqskip]}
\def    \ie             {{\em i.e.\/} }
\def    \eg             {{\em e.g.\/} }
\def\rd{{\mathrm d}}
\def\Im{{\mathrm{Im}}}
\def\mq{\mbox{$m_Q$}}
\def\mqq{\mbox{$m_{Q\bar Q}$}}
\def\mqqsq{\mbox{$m^2_{Q\bar Q}$}}
\def\kev{\mbox{$\mathrm{keV}$}}
\def\mev{\mbox{$\mathrm{MeV}$}}
\def\gev{\mbox{$\mathrm{GeV}$}}
\def\pt{\mbox{$p_T$}}
\def\ptsq{\mbox{$p^2_T$}}
\def\jpsi{\mbox{$J\!/\!\psi$}}
\def\chic{\mbox{$\chi_c$}}
\def\psp {\mbox{$\psi^{\prime}$}}
\def\ups {\mbox{$\Upsilon$}}
\def\mups {\mbox{$M_\Upsilon$}}
\def\height{\mbox{$H_8^{\prime}$}}
\def\vevpsi{\mbox{$\langle {\cal O}_8^{\psi}(^3S_1) \rangle$}}
\def\vevpsp{\mbox{$\langle {\cal O}_8^{\psp}(^3S_1) \rangle$}}
\def\as{\mbox{$\alpha_s$}}
\def\assq{\mbox{$\alpha_s^2$}}
\def\eps{\epsilon}
\def\lsim{\raisebox{-3pt}{$\>\stackrel{<}{\scriptstyle\sim}\>$}}
\def\bentarrow{\:\raisebox{1.1ex}{\rlap{$\vert$}}\!\rightarrow}
\def\dk#1#2#3{
\begin{array}{r c l}
#1 & \rightarrow & #2 \\
 & & \bentarrow #3
\end{array}
}
\newcommand{\bra}[1]{{\langle #1 |}}
\newcommand{\ket}[1]{{| #1 \rangle}}
\begin{titlepage}
\nopagebreak
{\flushright{
        \begin{minipage}{5cm}
        CERN-TH/95-190\\
	hep-ph/9507353 \\
        \end{minipage}        }

}
\vfill
\begin{center}
{\LARGE { \bf \sc Phenomenology of Quarkonium Production in Hadronic
Collisions} }  \footnote{To appear in the Proceedings of the Xth Topical
Workshop on Proton--Antiproton Collider Physics, Batavia, IL, USA, May 1995.}
\vfill
\vskip .5cm
{\bf Michelangelo L. MANGANO}
\footnote{On leave of absence from INFN, Pisa, Italy}
\vskip .3cm
{CERN, TH Division, Geneva, Switzerland} \\
\end{center}
\nopagebreak
\vfill
\begin{abstract}
We review recent progress made in the theory of quarkonium production in
hadronic collisions.
\end{abstract}
\vskip 1cm
CERN-TH/95-190\hfill \\
July 1995 \hfill
\vfill
\end{titlepage}

\section{Introduction}
The production of quarkonium states in high energy processes has recently
attracted a lot of theoretical and experimental interest.
The detection of \jpsi's plays a fundamental role in the study of $B$ physics,
because some of the most interesting final states of $B$ decays do contain a
\jpsi. It is therefore important to have a good understanding of all other
possible
sources of \jpsi's, in particular the direct production.
Furthermore, the large production cross sections and the relative ease
with which \jpsi's can be triggered on even at small values of transverse
momentum, make their observation a powerful tool to study hard phenomena
in regions of small $x$, which are otherwise inaccessible with the standard
hard probes (jets and vector bosons) used in high energy hadronic collisions.
Since we expect \jpsi's to be mostly produced via gluon-gluon fusion,
if a solid theory existed it could
be used to get the best and most direct measurements of the gluon density of
the proton at small $x$. The data are there, plentiful!

Production models have existed for several years (see
ref.~\cite{schuler} for a comprehensive review and references).
However, it is only with the
advent of the wealth of data from the high energy hadronic colliders
[2--6] that
significant tests of the theory have become possible, thanks to the big lever
arm in CM energy relative to the fixed-target experiments, and thanks to the
wide range in transverse momenta that can now be probed.
The comparison of these data with the models available up to a couple of years
ago has shown dramatic discrepancies, the most striking one (theory predicting
a factor of 50 fewer prompt \psp\ than measured by CDF
\cite{cdf94,mlm-glasgow}) having become known as the ``CDF anomaly''.

Attempts to explain the features of these data have led in the past couple of
years to a much deeper theoretical understanding of the mechanisms of
quarkonium production. None of these developments would have been possible
without the significant achievements of the experiments, to which most of the
credit should go.
In this review I summarize the evolution of the theoretical models towards
what we can consider today as the seed of a theory of quarkonium production
based on QCD. I will not have time to cover, however, the series of papers
\cite{close}\ that proposed the existence of new exotic charmonium states to
resolve the conflict between data and theory.

The language I use is inspired by a space-time picture of the production
process, and in my view should appeal to the generic reader for its simplicity.
It should be kept in mind, however, that most of the qualitative statements
that will be made can be rephrased in more rigorous terms, as discussed in the
references that will be quoted.

\section{Quarkonium Production for Pedestrians}
Production of quarkonium
represents a challenging theoretical problem, as it requires
some understanding of the non-perturbative dynamics responsible for the
formation of the bound state. The problem could be made easier by
assuming the existence of some factorization theorem that allows the separation
of the dynamics of the production of the heavy quark pair from its evolution
into a bound state. The reason why this assumption would make things simpler
is that we could,
in the presence of factorization, parametrize the
non-perturbative part in a universal fashion in terms of a limited set of
parameters: having been determined once (for example by fitting some set
of data), these can then be used to perform predictions.
The assumption of factorization is reasonable, since the time
scales associated to the two phenomena are significantly different: in the
production of the quark pair, the relevant time scale is the inverse of the
mass of the heavy quark, or of its transverse momentum in the case of high-\pt\
production. In the formation of the bound state, the important time scale is of
the order of the inverse of the quarkonium binding energy, \ie something of
the order of
$1/\Lambda_{QCD}$. Therefore, by the time the bound state starts forming, the
memory of what the source of the heavy quark pair was has been lost.
However, the quarkonium state has well defined quantum numbers, and one
might suspect that only heavy quark pairs {\em prepared} by the hard process in
specific states have a chance to eventually evolve into a given bound state.
Selection rules could therefore prevent the loss of memory, and spoil
factorization.

Determining to which extent, and in which precise form, factorization holds, is
therefore the primary challenge that we are faced with when formulating a
production model. I will now shortly describe the two most popular models which
have been proposed in the past, the so-called ``colour evaporation'' (CEM,
\cite{cem}) and ``colour singlet'' (CSM, \cite{csm}) models. They are based on
orthogonal assumptions about the validity of factorization and, needless to
say, lead to significantly different predictions. In a later section, I will
show how these models evolved in the recent past into a more sophisticated
approach, developed within QCD, which can apparently explain the main features
of the data currently available.

\subsection{Colour Evaporation}
In the colour evaporation (also known as local-duality) approach,
factorization is assumed to hold strictly. Differential distributions for the
production of a given quarkonium state $H$ are assumed to be proportional to
the production rate of a pair of heavy quarks with invariant mass in the range
$2\mq < \mqq < 2m_D$:
\be
	\frac{d\sigma(H)}{dX} \=  A_H
	\int_{2\mq}^{2m_D} \; d\mqq \, \frac{d^2 \sigma(Q,\bar Q)}{d\mqq dX}.
\ee
Here $Q$ is the heavy quark that forms the bound state, \mq\ is its
mass and $m_D$ is the mass of the lightest meson carrying open flavour $Q$,
\ie the $D$ meson in the case of charm, or the $B$ meson in the case of
bottom; \mqq\ is the invariant mass of the produced heavy quark pair.
The justification of this model stems from the assumption that only quark pairs
below the threshold for production of open flavour can possibly
bind into a quarkonium state, and that provided the quark pair has mass below
this threshold, the correct quantum numbers for $H$ will be recovered via
non-perturbative emission ({\em evaporation}) of very soft gluons. The constant
$A_H$, with $A_H<1$, depends on the state $H$ we are interested in, but is
otherwise independent of the \pt\ of $H$, and of the nature of beam and
target. Therefore, while the model cannot estimate absolute cross sections, it
however predicts their \pt\ and $\sqrt{s}$ dependence and it  predicts ratios
of  production rates and distributions of different states to be a constant.

As an example, consider production at large \pt: in this case
the dominant mechanism for the production of heavy quark pairs with invariant
mass close to threshold is the splitting of a high-\pt\ gluon: $g\to Q\bar Q$.
The probability for this splitting to take place can be calculated in leading
order (LO) QCD as:
\be
	\frac{d{\rm{Prob}}}{d\mqqsq} \=
	\frac{\as}{6\pi}\frac{1}{\mqqsq},
\ee
and therefore:
\ba
	\frac{d\sigma(H)}{dp_T^H} &=& A_H
	\int_{2\mq}^{2m_D} \; d\mqqsq \,
	\left ( \frac{d \sigma(g)}{dp_T^H} \right )
	\frac{\as}{6\pi}\frac{1}{\mqqsq} \nn             \\
	&=& A_H
	\frac{\as}{3\pi}
	\left ( \frac{d \sigma(g)}{dp_T^H} \right )
	\log \left( \frac{m_D}{m_Q} \right )
	\=
	A_H \frac{\as}{3\pi}
	\left ( \frac{d \sigma(g)}{dp_T^H} \right )
	\frac{2\epsilon}{m_H},
\ea
where we defined $\epsilon=m_D-m_Q \ll m_H$.

Attempts can be made to estimate what the relative values of $A_H$ for
different states $H$ should be. For
example, one could naively assume that a quark pair will evolve with a fixed
probability into the closest state $H$ with mass $m_H<\mqq$. In this case,
\be
	d\sigma(H) \propto (2J+1) \frac{\epsilon_H}{m_H},
\ee
where $\epsilon_H$ is the mass splitting between adjacent states. In the case
of charmonium, we would have for example $m_\chi-m_\psi \sim 300$ \mev,
$m_{\psp}-m_\chi \sim 200$ \mev, $2m_D-m_{\psp} \sim 100$~\mev,
so that:
\ba
\label{psppsi}
&&	\frac{\sigma(\psp)}{\sigma(\psi)} \sim 1/3  \\
\label{chichi}
&&  	\frac{\sigma(\chi_J)}{\sigma(\chi_{J'})} \sim \frac{2J+1}{2J'+1} \\
\label{chipsi}
&&  	\sum_J \sigma(\chi_J) \times BR(\chi_J \to \psi) / \sigma(\psi)
  	\sim 0.3.
\ea
It is interesting to compare these naive predictions with data. From the
publications of E705 \cite{e705} (300 GeV beams of pions or protons on
nuclei), the following ratios can be extracted:
\ba
&&	\frac{\sigma(\psp)}{\sigma(\psi)} \sim 0.25 \ret
&&	\sum_J \sigma(\chi_J) \times BR(\chi_J \to \psi) / \sigma(\psi)
  	\sim 0.6
\ea
The first result is also consistent with data from E789 \cite{e789}, if
the fraction of \jpsi's from
$\chi$ decays is assumed to be of the order of 0.3.

Results presented by D0 at this Conference indicate a fraction of \jpsi's from
$\chi$'s of approximately 0.4, however with a strong \pt\ dependence.
Likewise, CDF reported a value for this fraction of approximately 0.3, again
with some \pt\ dependence. CDF also measured the fraction of $\chi_1$
production relative to the total of $\chi_1+\chi_2$, of the order of 0.5. All
of these numbers come quite close to the naive expectations given in
eqs.~\eqnum{psppsi}--\eqnum{chipsi}.
The measurement of the total cross sections for $\Upsilon$ production by CDF
\cite{cdf95}\ has also been shown to be consistent with the $\sqrt{s}$
dependence predicted by the colour evaporation model \cite{schuler2}.

Therefore, while not completely satisfactory and not a real theory, it is
clear that the colour evaporation model presents some features of universality
that are consistent with the observed data, and that should therefore be
maintained by the final theory, however complicated this might be.

\subsection{The Colour Singlet Model}
The colour singlet model \cite{csm}, at least in its early formulation,
emphasizes more the constraints imposed by the colour and spin
selection rules. In the CSM, one projects the amplitude for the
production of a heavy quark pair directly onto
a state which has the right quantum numbers to form a given quarkonium
state. This projection singles out only the right combinations of colour and
spin required, and allows the absorption all the non-perturbative physics of
the confinement into a single parameter, namely the value of the wave function
(or derivatives thereof) of the quarkonium state at the origin.

For example, in the case of $^3S_1$ production this can be achieved
by evaluating the following expression \cite{csm}:
\be
	M(\psi(P)) \= \frac{R(0)}{\sqrt{16\pi m}}
	 \frac{\delta_{ij}}{3} \epsilon_\mu(P)
	{\rm Tr} [ {\cal O}^{ij} \gamma^\mu (\not\!\!P+m) ]
\ee
where $M(\psi)$ is the matrix element for the production of a $^3S_1$ state of
momentum $P$ and mass $m$,  $\epsilon_\mu$ is its polarization vector,  $R(0)$
is the wave function at the origin, and ${\cal O}^{ij}$ is the matrix element
for the production of the heavy flavour pair ($i$ and $j$ being the colour
indices of quark and antiquark), with the constraint that the relative momentum
between the two quarks be 0.
\begin{figure}[t]
\begin{center}
\psfig{figure=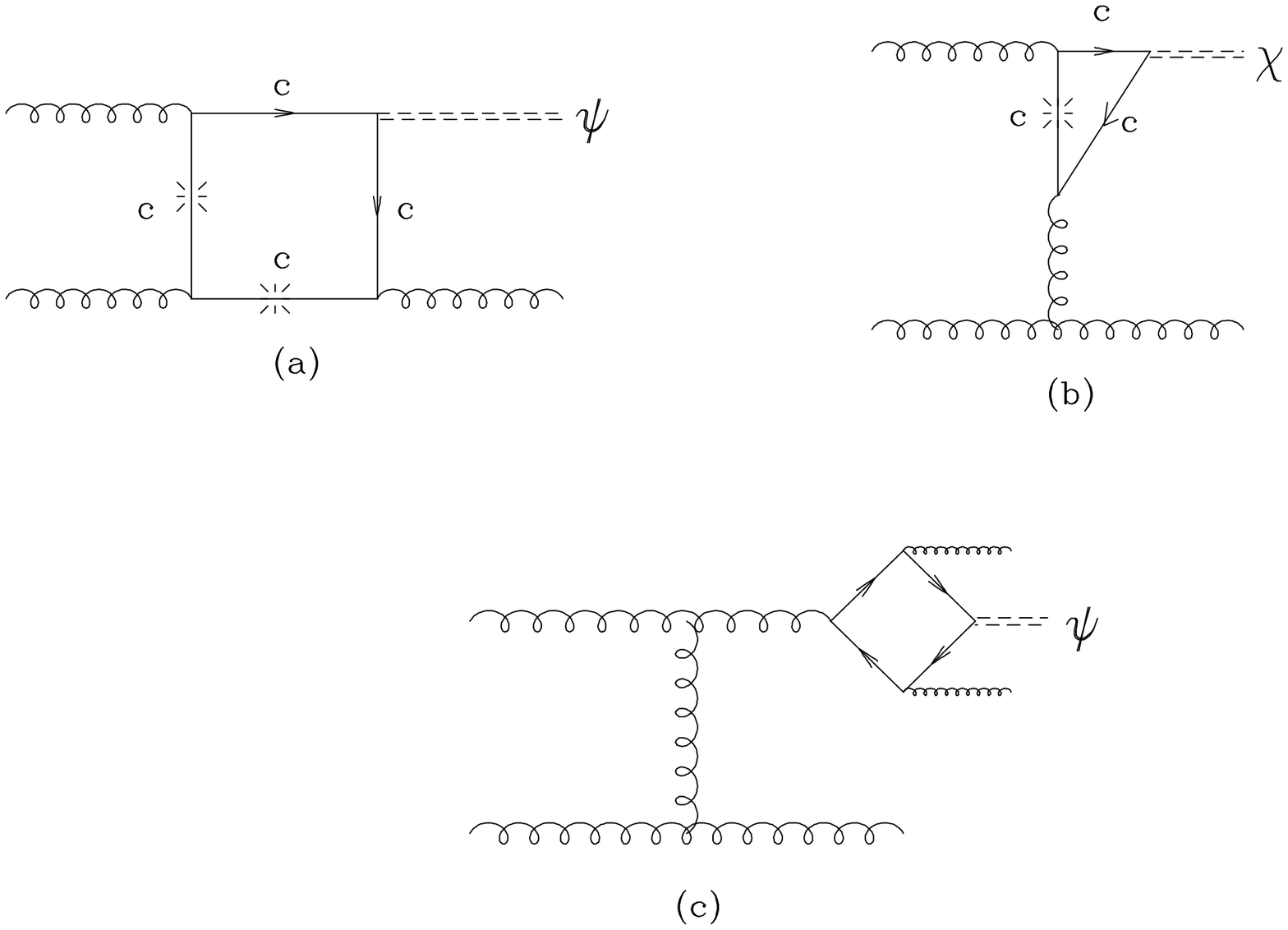,width=9cm,angle=90,clip=}
\ccaption{}{\label{fpsi}\small Sample graphs contributing to quarkonium
production
in the colour singlet model. (a): Leading order $^3S_1$.
(b): Leading order $^3P_J$.
(c): Fragmentation graph for $^3S_1$. The propagators crossed with a star
are those with $q^2 \sim -p_T^2$.}
\end{center}
\end{figure}
Similar projection operators can be evaluated for
any $^{2S+1}L_{J}$ state \cite{csm}, and convoluted with heavy quark production
matrix elements  calculable in QCD.  Absolute predictions can therefore be made
for the production rates, once we introduce the values of the wave functions
that can be extracted from potential models or directly from the data on
quarkonium decay widths.

In the case of hadronic collisions, and at the leading order in \as, namely
$\alpha_s^3$, it is easy to show that the only diagrams that are relevant for
the
production of a $^3S_1$ state are those shown in Fig~\ref{fpsi}a.  At large
transverse momentum, a simple kinematical analysis of the momentum flow in the
two internal quark propagators shows that they are off-shell by approximately
$q^2 = - (4 \mq^2 + p_T^2)$. As a result, the square of the matrix element will
behave at large \pt\ like $1/p_T^8$. This is a much steeper fall than
that of standard hard processes, such as jet production, where the behaviour is
that typical of a gluon or quark exchange, \ie $1/p_T^4$.
A similar analysis can be done in the case of large-\pt\ production of $\chi$
states, which is dominated by diagrams like the one shown in
Fig~\ref{fpsi}b. Here helicity conservation at the triple gluon vertex causes
the internal gluon propagator to behave like $1/\pt$, and the total amplitude
squared is therefore only suppressed by a factor of $1/p_T^6$. Therefore the
\pt\ distributions of \jpsi\ and $\chi$ as predicted at the LO by the CSM
are totally different, contrary to what is assumed in the CEM.

Such a steep \pt\ dependence has been proved to be inconsistent with the
Tevatron data, where the accessible range of \pt\ is very large \cite{cdf94}.
Aside from the technical details of how this behaviour arises from the
diagrams, there is a simple reason why the CSM predicts such a strong
suppression of high-\pt\ quarkonium production at LO.
In the CSM, one forces  the heavy quark pair to be in the right state
already at time scales significantly earlier than the time at which the
formation of the bound state starts. This requirement produces a strong
penalization in rate, which becomes more and more severe at higher \pt.
In fact at large \pt\ the time available for the pair to
organize itself into a state with the right quantum numbers becomes shorter,
and we pretend that it holds together, with nothing happening to it which could
change its state, until the exchange of Coulomb gluons takes over and binds it.
This phenomenon manifests itself with a strong form-factor-like suppression of
the production at large \pt, typical of such exclusive processes. Since the
probability that the quark pair can be found in the right state depends
directly on the details of the hard process which produced the pair, it should
come as no surprise that no universal factorization applies in this case,
and that the \pt\ slopes of \jpsi\ and $\chi$ differ.

It is possible \cite{bygluon}\ to incorporate within the CSM the effect of
longer time scales by considering higher-order contributions in perturbation
theory (PT). For example,
one could consider production of the heavy quark pair from gluon splitting at
large \pt\ (as in the case of the CEM), and describe the {\em perturbative}
evolution of this pair into a colour singlet state with the right quantum
numbers via gluon emission. The gluon virtuality before its splitting is of the
order of the quarkonium mass, and the time available for the quark
pair to evolve into the right state is larger than for the LO process.
Since we allow for the emission of gluons after the
creation of the pair, the process is no longer exclusive, but rather an
inclusive fragmentation process, and the form factor
suppression is avoided. In the case of \jpsi\ production, diagrams which
contribute to the production via fragmentation first appear at ${\cal
O}(\alpha_s^5)$ in PT (see fig.~\ref{fpsi}c).
They are therefore suppressed by a factor $\assq$ relative to
the LO contributions. However, their \pt\ behaviour is governed by the
exchange of the gluon in the $t$-channel, and is therefore given by
$1/p_T^4$.
The ratio of fragmentation
over LO cross sections is then of order $(\as \ptsq/\mqqsq)^2$. This becomes
larger than 1 as soon as \pt\ is larger than few times the quarkonium mass,
namely in the region where the Tevatron data come from.

\begin{figure}[t]
\begin{center}
\psfig{figure=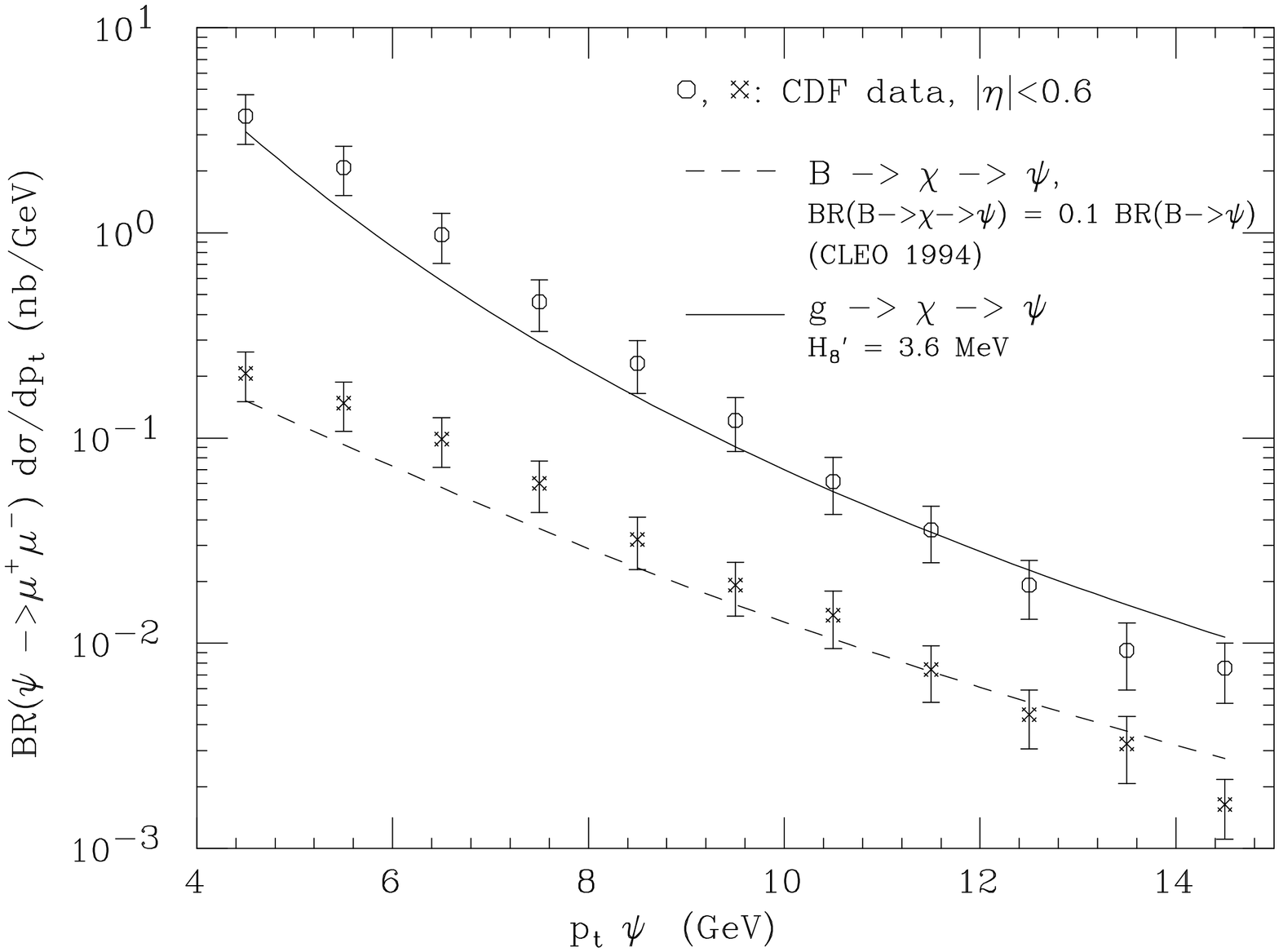,width=9cm,angle=90,clip=}
\ccaption{}{
\label{fchi}\small Inclusive $p_T$ distribution of $\psi$'s from
$\chi_c$ production and decay.
Upper curves and data points correspond to the prompt component.
Lower ones correspond to the $b$ decay contribution. CDF data versus theory.
The $b$ contributions were evaluated using the NLO matrix elements inclusive
$b$ production \cite{nde}\ and the CLEO measurements of $BR(b\to\chi$)
\cite{cleo}.
}
\end{center}
\end{figure}

The lesson to be learned is that in the case of quarkonium
production, naive \as\ power counting does not establish by itself the
correct perturbative expansion, in spite of the smallness of \as.
In fact, it turns out that the leading-order terms in \as\ represent
contributions that should be considered as {\em higher-twist} corrections.

\begin{figure}[t]
\begin{center}
\psfig{figure=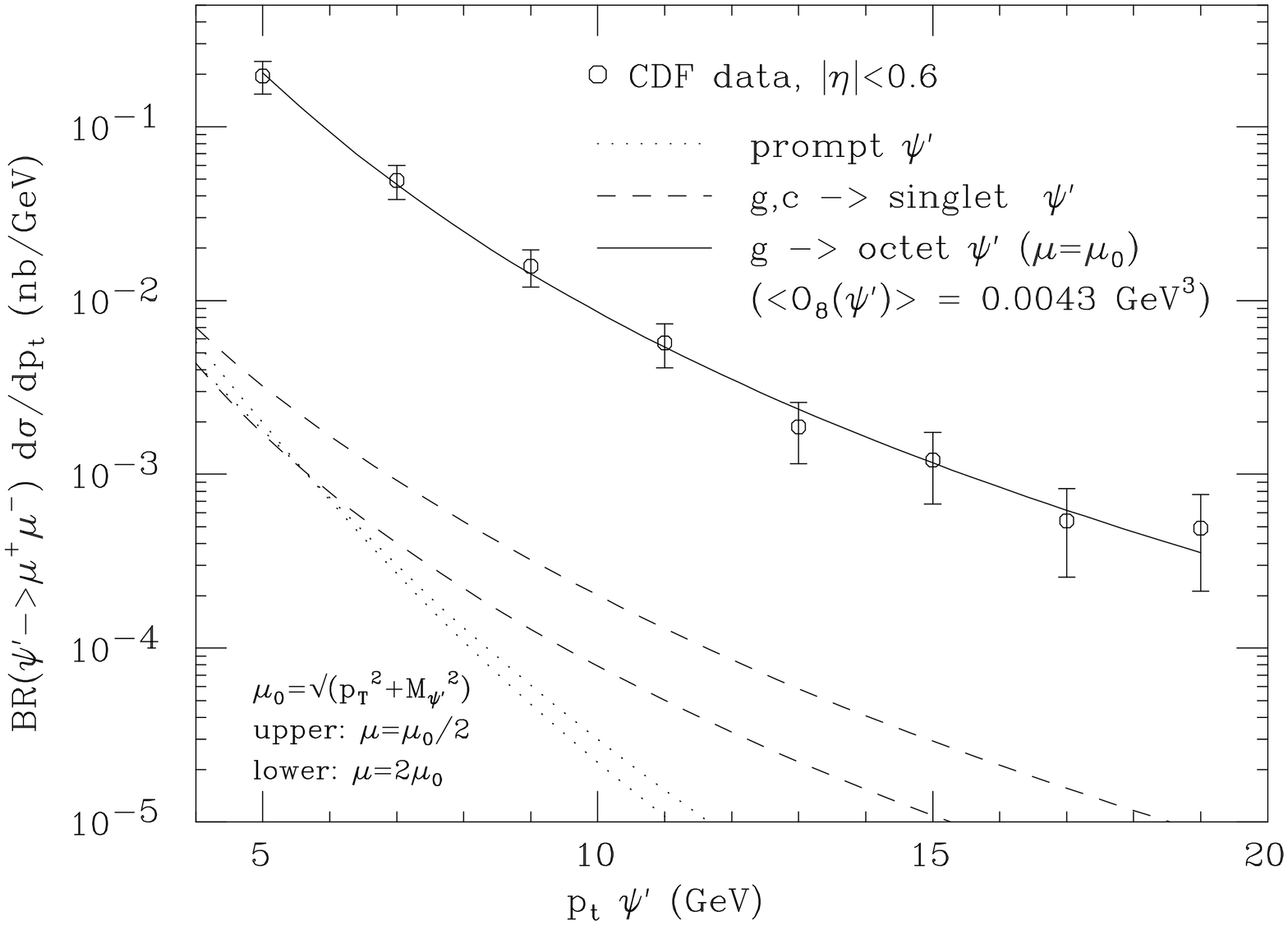,width=9cm,angle=90,clip=}
\ccaption{}{\label{fpspdata}\small Inclusive prompt $\psp$ $p_T$ distribution.
CDF data versus theory. We show the contribution of the different sources.
Dotted lines: LO production in the CSM; dashed lines: gluon and charm
fragmentation in the CSM; solid line: gluon fragmentation in the colour octet
mechanism.}
\end{center}
\end{figure}
The inclusion of fragmentation contributions bridges the gap between
the two extreme philosophies of the CEM and of the CSM: in the fragmentation
approach, factorization is achieved via the separation between the production
of a
hard (but almost on-shell, relative to the global $Q^2$ of the hard
process) gluon and the evolution of this gluon into a given
quarkonium state. This evolution is described by universal, although
state-dependent, fragmentation functions, which replace the simple-minded
overall constant $A_H$ introduced in the CEM.

Fragmentation functions for all states of interest have been calculated over
the past couple of years \cite{bygluon,frag}, and have been used for
phenomenological studies \cite{pheno}. In the case of $\chi$ production, the
theoretical calculations agree with the available data, as shown in
fig.~\ref{fchi}\ \cite{cgmp}. In the case of the $^3S_1$ states, however, the
discrepancy in overall normalization is striking, although the \pt\
distribution fits the data well (see dotted and dashed curves in
fig.~\ref{fpspdata}).
The conclusion is that while fragmentation
contributions are fundamental to produce the right \pt\ dependence, and are
sufficient to correctly predict the absolute normalization of the $\chi$ cross
section, there must be additional contributions to explain the abundance
of \psp\ produced.

\subsection{The Colour Octet Mechanism}
In order to understand the possible origin of these residual discrepancies, one
has to look more closely into the structure of the fragmentation process within
the CSM. In the case of fragmentation into a $^3S_1$ state, the
transition of a gluon into the $J^{PC}=1^{+-}$ state and only one gluon is
forbidden. Therefore emission of at least two gluons is required
(fig.~\ref{fpsi}c).
It turns out from the explicit
calculation of the fragmentation probability that only hard gluons contribute
to this process: soft gluons occupy a small volume of phase space,
and there is no dynamical enhancement in their emission. Therefore the
fragmentation probability is proportional to $\alpha_s^3(m)$. On the contrary,
the fragmentation of a gluon into a $\chi$ state requires emission of just one
gluon, as allowed by the different quantum numbers of the $^3P_J$ states.
Furthermore, emission of a soft gluon is enhanced by the presence of a well
known logarithmic infrared singularity \cite{barbieri}, which can be regulated
by noticing that the heavy quarks inside the bound state are slightly
off-shell, therefore cutting off the singularity at an energy of the order of
few hundred \mev. So the probability for a gluon to evolve into a $\chi$
state is of the order of $\as(m)$, as the large logarithm compensates the
additional power of $\as$.  As a net result, production of $\chi$'s is
significantly enhanced relative to that of \jpsi's.

Again one can interpret this phenomenon using time-scale arguments. Since the
evolution of a gluon into a \jpsi\ in the CSM only involves the emission of
hard gluons, the time scale for the transition is of the order of their
energy in the virtual gluon rest frame. In the case of the $\chi$,
instead, the enhancement of soft gluon emission indicates that the lifetime of
the heavy quark pair resulting from the gluon splitting, before it settles into
the $\chi$ state, can be very long.
Bodwin, Braaten and Lepage developed recently a framework \cite{bbl}, based on
non-relativistic QCD, in which such a long-lived state has a non-zero overlap
with the $\chi$. In their formulation, a quarkonium wave function
is the sum of contributions coming from states in the Fock space in which the
heavy quark pair is accompanied by long-lived gluons. In the specific case of a
$\chi$, one has:
\be
	|\chi_J\rangle = O(1) | Q \bar Q[^3P_J^{(1)}] \rangle +
	O(v) | Q \bar Q[^3S_1^{(8)}] g \rangle + \dots
\ee
where the upper indices $(1)$ or $(8)$ refer to the colour state of the pair,
and $v$ is the velocity of the heavy quark in the bound state.
The first term in the expansion corresponds to the standard non-relativistic
limit, in which the quarkonium is made just of the quark-antiquark pair. The
second term corresponds to a state in which the pair, in a colour octet
configuration, is accompanied by a gluon (the angular momentum of the quark
pair is different in the second state, as the gluon itself carries spin). It is
precisely this second component of the $\chi$ state that has non-zero overlap
with the long-lived pair coming from gluon splitting. Its presence allows
the infrared divergence alluded to above to be absorbed via a wave function
renormalization, in a way which can be rigorously defined and extended to
higher orders in PT, and which does not have to rely on an arbitrary IR cutoff.
The final picture we obtain is therefore as follows:
the fragmentation function for production of a $\chi$ state from gluon
evolution consists of two pieces. One is of order $\assq$, and corresponds to
the creation of a heavy quark pair with the emission of a perturbative gluon,
the quark pair being projected on a short time scale on the $^3P_J^{(1)}$
state. The other is of order $\as v^2$, and corresponds to the creation of a
quark pair in the $^3S_1^{(8)}$ state. This state will evolve on a long time
scale into the required $^3P_J^{(1)}$ state via emission of a non-perturbative
gluon.
The separation of short time scales from long ones is arbitrary, but the
result is independent of it, as its redefinition only leads to a change in the
relative importance of the two contributions. The production of $\chi$ through
this second component has been named ``colour octet mechanism'' (COM). Since it
turns out that, for charmonium, $\as$ is numerically
of the same order as $v^2$, the two channels are competitive.

Braaten and Fleming \cite{bf}\ suggested that a similar phenomenon might play a
key role in \jpsi\ production as well. Colour octet states do in fact
appear in the expansion of the $^3S_1$ state \cite{bbl}:
\ba
	|\psi\rangle &=& O(1) | Q \bar Q[^3S_1^{(1)}] \rangle +
	O(v) | Q \bar Q[^3P_J^{(8)}] g \rangle \nn \\
	&+&
	O(v^2) \left ( \;
	| Q \bar Q[^3S_1^{(1,8)}] g g \rangle +
	| Q \bar Q[^1S_0^{(8)}] g \rangle +
	| Q \bar Q[^3D_J^{(1,8)}] g g \rangle \; \right ) + \dots
\label{psifock}
\ea
There is a state in this decomposition which can be accessed by a gluon
already at order \as, namely
$| Q \bar Q[^3S_1^{(8)}] g g \rangle$.
Creation of this state via gluon fragmentation will be suppressed by a factor
$v^4$, but this can be compensated by the absence of the two extra powers
of \as\ that  are required for production of the colour singlet state. So once
again the two processes could be competitive. When one performs the
complete calculation, numerical factors appear which leave the COM as by far
the dominant one. If we take for example the integral of the fragmentation
functions \cite{bygluon,bf}, we obtain the following ratio of probabilities:
\be
	\frac{[{\rm Prob}(g\to {^3S_1})]_{octet}}{[{\rm Prob}(g\to
	{^3S_1})]_{singlet}} \approx
	25 \frac{\pi^2}{\assq(m_\psi)}
	\frac{\langle O_{^3S_1}^{(8)} \rangle}{\langle O_{^3S_1}^{(1)} \rangle},
\ee
where the two objects in the last ratio are defined in \cite{bbl}\ and
can be shown to be proportional to the square of the wave functions at the
origin for the colour octet and colour singlet components of the \jpsi\ state,
respectively. The factor $\pi^2$ arises as a simple consequence of the phase
space for the two additional gluons present in the final state of the colour
singlet fragmentation. In order for the octet contribution to be larger than
the colour singlet one by a factor of 50 (the CDF \psp\ anomaly),
it is therefore sufficient that
\be \label{vev}
	\langle O_{^3S_1}^{(8)}  \rangle \sim
	\frac{2}{\pi^2} v^4 \langle O_{^3S_1}^{(1)}  \rangle,
\ee
where we used
the fact that numerically $\as \sim v^2$. This is
exactly of the right order of magnitude implied by
eq.~\eqnum{psifock}.

\section{Comparison with the Tevatron and
$ S\lowercase{p\bar p}S $ data}
\begin{figure}[t]
\begin{center}
\psfig{figure=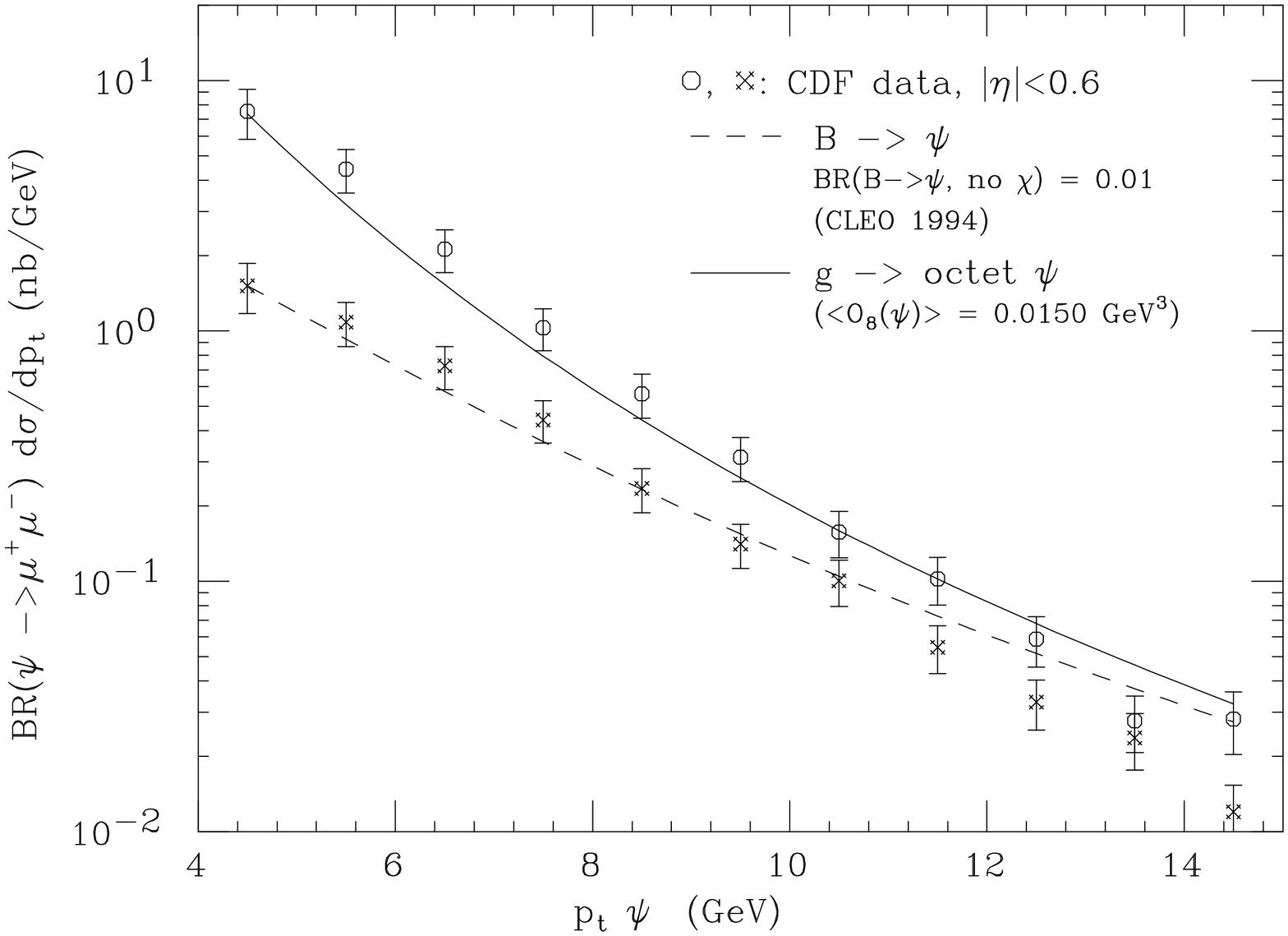,width=9cm,angle=90,clip=}
\ccaption{}{
\label{fpsidata}\small Inclusive $\psi$ $p_T$ distribution.
Upper curves and data points correspond to prompt $\psi$'s, after subtraction
of the $\chi_c$ contribution. Lower ones correspond to the $b$ decay
contribution.
CDF data versus theory.}
\end{center}
\end{figure}
The introduction of the COM provides a potentially important new production
channel that could explain the CDF \psp\ anomaly. Unfortunately, aside from
the generic feature that $\langle O_\psi^8 \rangle $ should be of order $v^4$
relative to $\langle O_\psi^1 \rangle$, we have no precise estimate of its real
value, although it is expected \cite{bbl}\ that lattice calculations could
provide it one day. What can be done, therefore, is to extract $\langle
O_\psi^8 \rangle $  directly from the data, fitting the CDF measurement to the
theoretical \pt\ distribution of \jpsi\ and \psp\ \cite{cho,cgmp}.

The results of the
fits are shown in fig.~\ref{fpspdata}\ and \ref{fpsidata}. As had already been
shown by Braaten and Fleming, the predicted shapes agree very well in the case
of the \psp. Now that data are also available \cite{cdf95}\
for the production of prompt
\jpsi's (\ie \jpsi's not coming from either $b$ or $\chi$ decays), we can reach
the same conclusion also for the lowest-lying $^3S_1$ state. Had we only
included the prediction of the CSM, the disagreement with the \jpsi\ data would
have been again of the order of 50. The extracted
values for the two new parameters $\langle O_{\psi}^8 \rangle $ and
$\langle O_{\psi'}^8 \rangle $ are given in the figures, and it can be verified
that they agree with the crude estimate given in eq.~\eqnum{vev}.

\begin{figure}[t]
\begin{center}
\psfig{figure=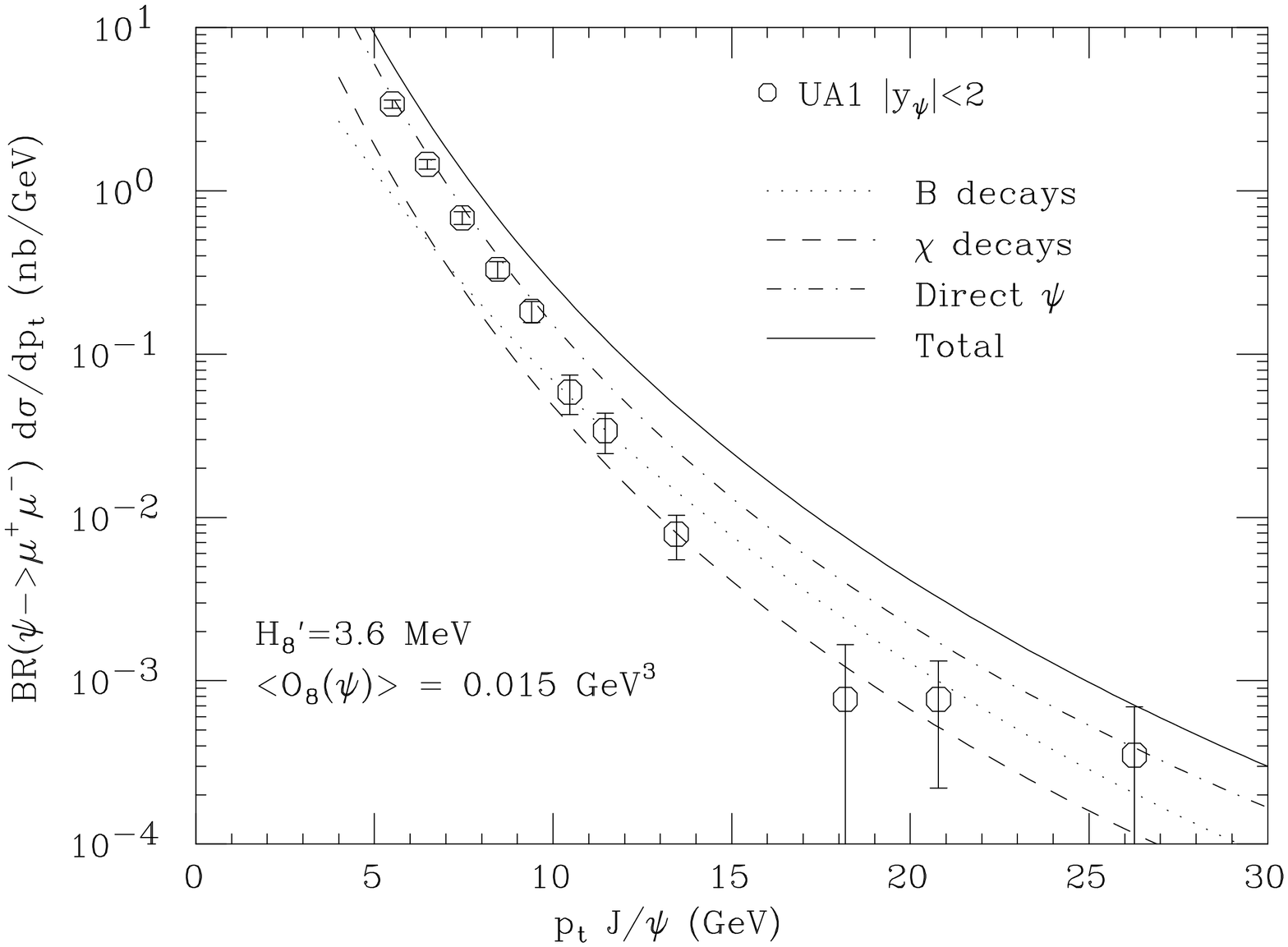,width=9cm,angle=90,clip=}
\ccaption{}{
\label{fua1}\small Inclusive $p_T$ distribution of $\psi$'s at 630
GeV. All sources of $\psi$ production are here included.
UA1 data versus theory. The parameters of the
theoretical calculation take the values fitted on the Tevatron data. }
\end{center}
\end{figure}

Having determined the values of these unknown parameters, one can use them
to predict rates in different experimental set-ups. In fig.~\ref{fua1}, for
example, we show the prediction for \jpsi\ production in $p\bar p$ collisions
at 630 \gev, compared to the UA1 data \cite{ua1psi}. In order to match the
experimental analysis, which did not separate between different sources of
$\psi$ production, we included the contribution of $b$ and $\chi$ decays as
well. The theory is higher by a factor of 2. It is possible that small-$x$
effects \cite{smallx}, which are expected to be responsible
for the increase of the $b$ production rate at
the Tevatron w.r.t. to the NLO calculation by an additional
factor of 30\% relative to 630 \gev\ \cite{lathuile}, play an even more
important role in the case of charm. This would cause our fit to the Tevatron
\jpsi\ and \psp\ data to overestimate the values of the parameters.

A comparison with fixed-target data, where the \pt\ values accessible are much
smaller, will require additional work, in order to absorb into the NLO parton
densities the collinear divergences which appear al low \pt\
\cite{inprogress}. Furthermore, at these low \pt's one will need to take into
account the effects due to the intrinsic Fermi motion of initial-state gluons
in the proton \cite{e789}.

Another important test of the COM comes from the study of $\Upsilon$
production. Here the colour octet effects are however
expected to be smaller, as $v$ is smaller. Nevertheless the complete
calculation of $\Upsilon$ rates within the colour singlet model predicts cross
sections which are smaller than data \cite{cdf95,d095}\
by a factor of 3 in the case of 1S and 2S
states, and by a factor of 9 in the case of the 3S.
While the factor of 3 discrepancy between the CSM prediction and the data could
be explained by the addition of the COM contributions, the
relative factor of 3 discrepancy between the
rates of 3S and 1S and 2S states is puzzling.
It is very tempting to assume that a yet unseen third set of $\chi_b$ states,
$\chi_b^{\prime\prime}$, exists below threshold, and decays radiatively to the
3S state. Various potential model calculations of the bottomonium spectrum
support this idea.  Simple-minded estimates of the $\chi_b^{\prime\prime}$
contributions show that this process can correct the 3S yield by the required
factor of 3.

\begin{figure}[t]
\begin{center}
\psfig{figure=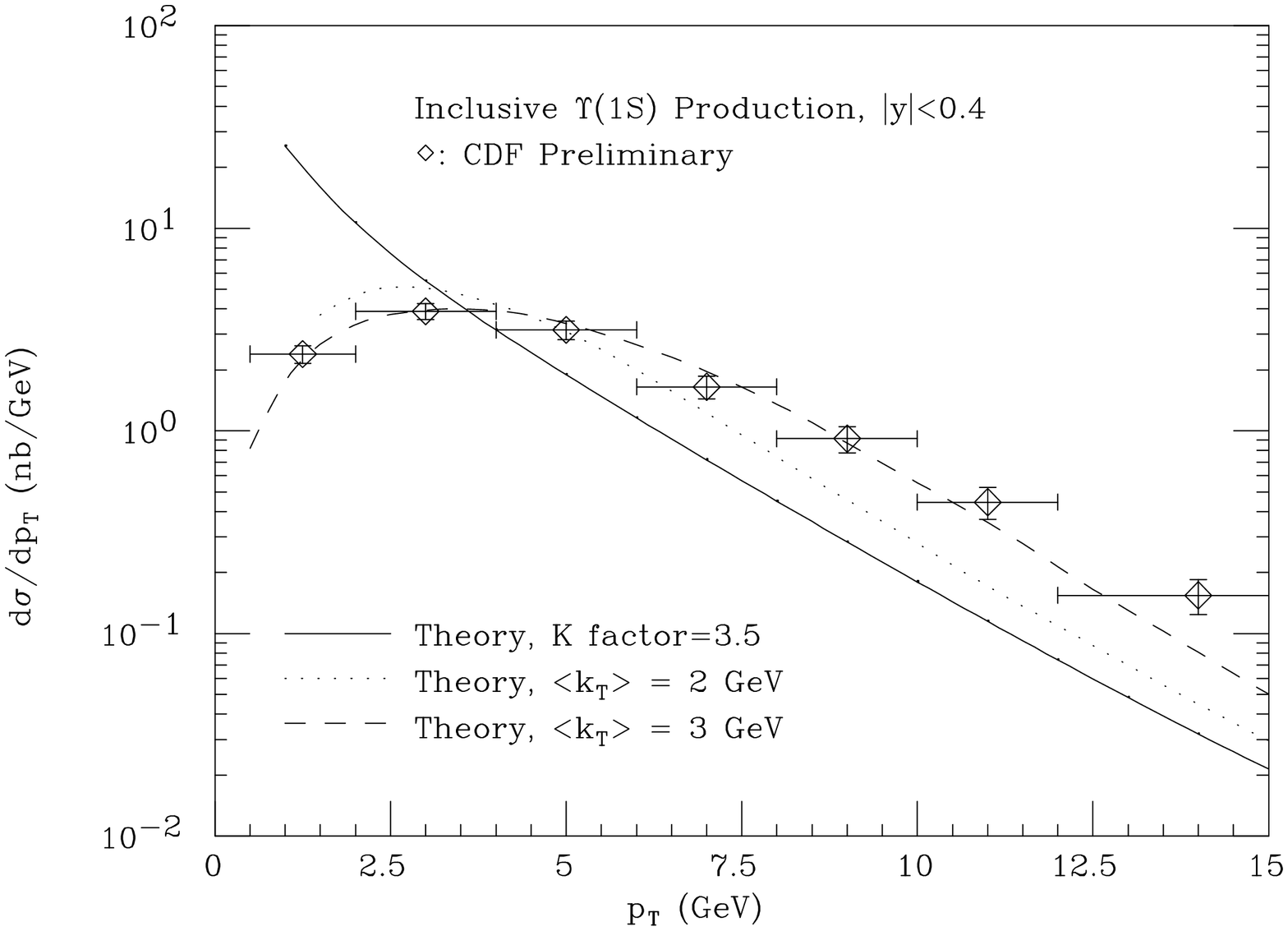,width=9cm,angle=90,clip=}
\ccaption{}{
\label{fups}\small Inclusive $p_T$ distribution of $\Upsilon(1S)$'s at the
Tevatron. Shown are data from CDF and the prediction of the CSM, rescaled by a
factor of 3.5. The three curves correspond to the inclusion of a $k_T$ kick of
0, 2 and 3 \gev.}
\end{center}
\end{figure}

The shapes of the \pt\ distributions are in principle sensitive to the
resummation of multiple soft gluon emission from the initial state, as the \pt\
values probed by the experiments are small relative to the mass of the
$\Upsilon$. Figure~\ref{fups}\ shows for example the effect of an additional
$k_T$
kick of the order of 2--3 GeV on the LO distributions, compared to the
$\Upsilon(1S)$ CDF data. The comparison with the D0 data is similar, both in
rate and shape.

Shortly after this Conference, Cho and Leibovich \cite{cho}\
presented a full calculation
of the contribution of the COM to the $\Upsilon$ rates, extrapolating to the
$b\bar b$
system the values of the non-perturbative parameters obtained from the fits to
the charmonium data. To properly describe the region $\pt<\mups$,
where most
of the data come from and where the use of fragmentation functions is not
appropriate, these authors performed the complete calculation of the LO Feynman
diagrams producing the $^3S_1^{(8)}$ state.
Their results indicate good agreement with the data,
for the 1S and 2S states. In the case of the 3S state, a residual factor of 3
discrepancy remains if one only includes the decays of the known $\chi_b$ and
$\chi_b^{\prime}$ states.
Their detailed estimate of the effect of  production and decay of
$\chi_b^{\prime\prime}$ states confirms nevertheless the view that their
existence would solve even this residual problem.
Although they are hard to detect directly, it is not unlikely that future
larger statistics accumulated at the Tevatron will allow their unambiguous
discovery.

\section{Conclusions}
The field of heavy quarkonium production has enormously benefited from the
recent measurements performed at the Tevatron Collider. Very rarely in the
recent past have experimental data been so important in guiding the development
of a theory. The initial discrepancies by almost two orders of magnitude found
between earlier models and the data have driven theorists to deepen their
understanding of the underlying dynamics, and have eventually led to a solid
framework within which to operate. The inclusion of the colour octet mechanism
is now viewed as a necessity for the consistency of the theory, rather than as
an {\em ad hoc} theoretical concoction. While the field is still in its
infancy, and additional progress must be made before a complete theory  is
formulated and firmer predictions can be made, I believe one can conclude that
we are on the right track. Several calculations are in progress, which will
eventually allow us to test the theory more thoroughly.
\\[0.3cm]
{\bf Acknowledgements:} I wish to thank the people who shared with me their
insight on this subject, and with whom most of the work described here was
carried out: E. Braaten, P. Cho, M. Cacciari, M. Doncheski, S. Fleming, M.
Greco and A. Petrelli.

\end{document}